\documentstyle[11pt,newpasp,twoside,epsf]{article}
\begin{document}

\title{The host galaxy - AGN connection at low and high redshift.}
\author{Renato Falomo}
\affil{INAF - Osservatorio Astronomico di Padova vicolo Osservatorio 5 \\
35122-Padova, Italy}

\begin{abstract}
The properties of radio loud active galaxies 
(radiogalaxies, BL Lac objects and radio loud quasar) at 
low and high redshift are briefly reviewed and 
compared. The recently derived empirical relations between central 
black hole mass and host properties allow one to explore the demographics 
of massive black hole in active galaxies. 
It is shown that all classes of active galaxies 
considered exhibit very similar host properties and, consequently, 
have central black holes of comparable mass. The  future observing 
capabilities would allow the investigation of the evolution of the host galaxy 
properties up to z $\sim$ 3, and will give insight into the joint 
formation of black holes and massive spheroids.
\end{abstract}

\section{Introduction}

The discovery that many and probably  all apparently 
inactive nearby early type galaxies harbour a dormant 
black hole (BH) in their nuclei (e.g. Ferrarese 2002) has 
changed the view that distinguished active from inactive galaxies. 
It is also becoming clear that a full comprehension of the processes 
that produced the galaxies we observe today must take into account 
the formation of supermassive black holes in their centers.

It is well assessed that the high energy phenomena observed in AGN 
occur in the nuclei of massive galaxies. Less understood is, however, the link 
between the properties of the active nucleus and those of the host galaxy. 
Nevertheless a number of issues are emerging from recent studies.
Nuclear activity associated with radio emission is present almost 
exclusively on galaxies dominated by the spheroidal component.
On the other hand radio quiet objects are found in both types of galaxies 
(bulge and disc dominated galaxies). 

A detailed comparison of the host properties of various 
type of AGNs together with  those of normal (inactive) galaxies 
can help to investigate the origin of the nuclear activity. 
In particular one can explore if and how the nuclear activity depends on 
the global properties (e.g. total luminosity and scale-length)
of the galaxies, if there are difference in the 
stellar population with respect to inactive galaxies 
and what is the role of interactions (disturbed morphologies, close companions).  
Such kind of comparison can be done with different degree of details that depends 
in part on the distance of the sources and on the prominence of the nucleus 
with respect to the host galaxy. Radiogalaxies are the nearest active objects 
and exhibit only faint nuclei and are therefore the most easy AGN to study. 
On the other hand quasars are rare objects locally, have extremely bright 
nuclei and are rather more difficult to study.

The recent discovery that the BH mass (M$_{BH}$)
is correlated with the properties of the bulge component of the 
host galaxy, which is translated into the relationships between M$_{BH}$ and the 
bulge luminosity 
(Kormendy \& Richstone 1995; Magorrian et al. 1998; 
Richstone et al. 1998; Kormendy \& Gebhardt 2001) and between
M$_{BH}$ and the velocity dispersion $\sigma$ of the host galaxy 
(Ferrarese \& Merritt 2000; Gebhardt et al. 2000;  
Merritt \& Ferrarese 2001b) offers a new tool for evaluation of 
BH masses in AGNs if a reliable measurement of 
host galaxy luminosity or velocity dispersion is done. 

While for AGN with strong emission lines (as QSO and Seyfert galaxies) 
the standard methods 
(e.g. reverberation mapping) under virial assumptions 
of the emitting regions can be  used to derive M$_{BH}$ 
( see e.g. Wandel et al. 1999 and Kaspi et al. 2000), the above relations 
may be  the only way to estimate M$_{BH}$ for active galaxies 
that lack of emission lines (as BL Lac objects) or that are too far (most of  
nearby radiogalaxies) 
to resolve the region of influence of the BH.

In addition the knowledge of 
the properties of the galaxies hosting active nuclei may also yield fundamental 
insight for probing the unification models of AGN.
The main ingredient of the unification scenario of radio loud AGN (e.g. 
Urry and Padovani 1995) is that the 
observed properties of an object may depend on orientation effects 
(because of an isotropic obscuration or emission from the nucleus). 
Two objects with identical intrinsic properties may therefore exhibit 
apparent different nuclear activity because they are seen at different angles. 

A simple test for this hypothesis is  to compare the properties 
of the host galaxies (that do not depend on orientation) of various classes 
of AGN. Radio loud AGNs constitute only a small fraction of active galaxies 
but they seem to form a  homogeneous class of sources whose 
phenomenology could be explained within the same scenario.

In this paper I'll first review the properties of the galaxies 
hosting low redshift (z$<$ 0.5) radio loud active nuclei.
This includes radiogalaxies, BL Lac objects and radio loud quasars (RLQs).
Then using the relationships between BH mass and host properties 
I  derive and compare the distribution of BH masses in the various 
classes of active galaxies. 
In the second part of this work I'll 
sketch the view of quasar hosts beyond z $\sim$ 1. 
Finally a summary of the main conclusions of this work are given  
together with a perspective of the future studies on high redshift sources.
\section{Host galaxies of radio loud active nuclei at z $<$ 0.5.}
In this section I'll briefly review the properties of the galaxies hosting 
active nuclei that exhibit strong radio emission.
These include radio galaxies at z $<$ 0.2, BL Lac objects and radio loud quasars at z $<$ 0.5. In order to make the comparison of host properties 
as much as possible homogeneous I 
considered ground based data for low redshift radiogalaxies while 
only observations taken with HST were considered for the active galaxies 
with bright nuclei 
(BLL and RLQ). Moreover all the data presented here have been 
homogenized in terms of used passband, galactic extinction, k-correction 
and use the same cosmological parameters 
( H$_0$ = 50 km s$^{-1}$ Mpc$^{-1}$ and
deceleration parameter q$_0$ = 0 are used throughout this paper).
\subsection{Radiogalaxies}
Nearby radiogalaxies (RGs) are the most easy active objects 
to study because of they are relatively abundant in the local universe and 
their nuclei are faint compared with the starlight emission.
 The largest and homogeneous optical study of nearby 
radiogalaxies has been presented by Govoni et al (2000). 
They report detailed surface photometry for 79 objects at z $<$ 0.12 
and extracted from two complete samples of radiosources 
(see Fasano et al 1996 for details).
It turns out that RGs  are systems dominated but the spheroidal 
component with average absolute magnitude 
$\langle M_{R}(tot)\rangle=-24.0$; FRI sources are
hosted in galaxies $\sim$ 0.5 mag more luminous than 
FRII galaxies.

The detailed study of the luminosity profiles 
showed  the presence of nuclear point sources whose luminosity is about few
percent that of the whole galaxy. The luminosity of this component
appears correlated with the core radio power but independent of the
luminosity of the host galaxy. This result has been well confirmed by 
HST observations where these nuclei are well visible on direct images 
(Chiaberge Capetti \& Celotti 1999).
The shape of the luminosity profile has also emphasized the 
frequent presence of an excess of light with respect to the bulge component 
which could be interpreted as due to a conspicuous halo.
As far as the internal structure of these galaxies is concerned (ellipticity, 
isophote shape, twisting), it was shown that 
radio galaxies are indistinguishable from the normal (non radio) ellipticals.

As a whole, these studies have therefore shown that the optical 
morphological/structural and photometrical properties of
radio and non-radio ellipticals are remarkably similar, suggesting 
that all ellipticals may go through a phase of nuclear activity
lasting for a small fraction of the total life of the galaxy.  

At variance with this, however, it was found that RG are on average slightly 
bluer and with a bluer gradient across the galaxy with respect to a sample of 
non radio ellipticals. 
This suggests a  possible effect connecting the phase of radio activity with that of star formation. 

It is known that the global properties of early--type galaxies
are fairly well described through a three dimensional space of
observables which, besides the effective radius $r_e$ and the
corresponding average surface brightness $\mu$, involves the central
velocity dispersion $\sigma_c$ (Djorgovski \& Davis 1987, Dressler et al
~1987). A full comparison between radio and non radio galaxies needs therefore 
to take into account also the kinematical properties.

Using photometrical and dynamical data for 73 
low red-shift (z$<$0.2) radio galaxies (Bettoni et al 2001) were able to 
compare the Fundamental Plane (FP) of RG (see Fig 1) with that defined 
by inactive ellipticals (J\o rgensen et al. 1996, JFK96). They showed that 
the same FP holds for both radio and non radio ellipticals. Radio galaxies 
occupy the region of the most luminous and large ellipticals.
The consistency even of the kinematics properties 
lead further support to the idea that virtually, all
ellipticals have the basic ingredients for becoming active.

\begin{figure}[ht]
\plotone{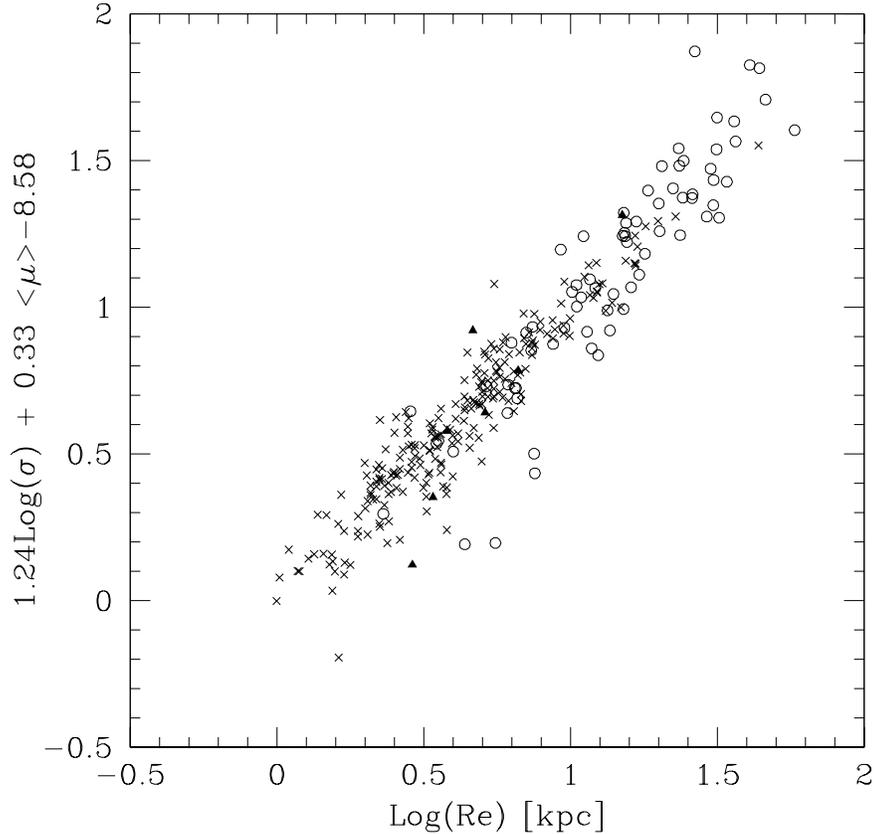}
\caption{The Fundamental Plane of the low red-shift 
Radio Galaxies ({\it open circles} ; Bettoni et al 2001) 
compared with the one ({\it crosses}) or non radio ellipticals (JFK96) 
and the host galaxies of 7 BL Lac objects ({\it filled triangles}; Falomo et al 2002). }
\end{figure}

\subsection{BL Lac objects}
Until about a decade ago the host galaxies of BL Lacs were poorly investigated 
and only for the nearest and most famous sources the 
global properties of the galaxies were known (see e.g. Ulrich 1989).
The dramatic improvement of the  telescope instrumentation and detectors 
has allowed one to perform a more systematic  study of 
the faint nebulosity associated with BL Lacs. Using high quality imaging 
obtained from ground based telescopes (mainly CFHT at Hawaii, NTT at ESO Chile 
and NOT at La Palma) various 
groups have provided systematic measurements of the basic properties of the hosts (Wurtz et al 1996 Falomo 1996, Falomo and Kotilainen 1999, 
Nilsson et al 2002 this conference) for large samples of objects.
These works have consistently shown that BL Lac hosts are virtually 
all massive ellipticals  
(average luminosity M$_R$ = -23.7 and effective radius Re = 10 kpc)  located 
in poor groups (but there are some notable exceptions: 
PKS 0548-32, Falomo et al 1995)
The few exceptions to this scheme (i.e. claims of disc dominated host galaxies) 
regards specific  objects (e.g. AO 0235+164, PKS0537-441, MS0205.7+3509) 
that were proposed to be candidates for microlensing 
(e.g. Stickel et al 1988a,b;
Stocke Wurtz \& Perlman 1995)
For these sources various pro and con observations have been 
reported and discussed in the literature (Abraham et al 1993; 
Falomo Melnick \& Tanzi 1992; Falomo et al 1997; 
see also Heidt at this conference). 
To the present time there is not convincing case
of a BL Lac object hosted in (or superposed to) a disc dominated galaxy.  

A further significant improvement to the knowledge of the properties 
of the galaxies hosting BL Lacs has been provided by the R band 
( F702W filter) high resolution images collected 
by WFPC2 camera on-board of HST during a snapshot (short exposure) 
survey of BL Lacs (Scarpa et al 2000, Urry et al 2000). This has produced 
a homogeneous set of high quality images for 110 BLL from which their 
environments can be investigated.
All (57) but three  targets at z $<$ 0.5 have been well resolved 
and emphasized the early type morphology of the hosts. 
Characterization of the host global properties indicated that they 
are indistinguishable from the population of luminous 
inactive ellipticals.  These data  also showed that 
no significant difference is present between the hosts of HBL  and 
those of LBL in spite of the remarkable difference of their SED 
(Urry et al 2000). 

Moreover for lower redshift (z $<$ 0.2) objects  a detailed study of 
isophote centering, twisting and isophote shape was performed 
(Falomo et al 2000).
It was found that both the ellipticity and the isophotal shape distributions 
are similar to those of radio galaxies and radio-quiet ellipticals. This
suggests that tidal interactions are very infrequent or are
short-lived with respect to the nuclear activity time scale.
Moreover no indication of  off-centering of the galaxy
isophotes with respect to the nucleus (with accuracy of $\sim 0.05$~arcsec) was 
found, meaning that the unresolved nuclear
source truly sits at the center of its galaxy.
This rules out the microlensing hypothesis for BL Lacs, which predicts frequent
off-centering of the nucleus (Ostriker \& Vietri 1990).
HST imaging therefore confirm with high level of details that the 
hosts of BL Lacs are not different from "normal" 
unperturbed massive spheroidal galaxies.

\subsection{Radio loud quasars}
The host galaxies of low redshift (z $<$ 0.5) quasars have been 
thoroughly investigated using both ground-based imaging (e.g. 
McLeod \& Rieke 1995; Taylor et al. 1996; Percival et al. 2000) and the 
Hubble Space Telescope (HST; e.g. Bahcall et al. 1997; Boyce et al. 1998; 
McLure et al. 1999). These investigations have shown that most quasars live 
in galaxies at least as bright as the Shechter function's characteristic 
luminosity L$^*$ (e.g.  Mobasher, Sharples \& Ellis 1993). While most quasar 
host galaxies are brighter than L$^*$, and in many cases comparable to 
brightest cluster member galaxies (BCM; Thuan \& Puschell 1989), some 
undetected or marginally detected hosts may be under-luminous. Recent 
morphological studies of host properties at low $z$ (e.g. Taylor et al. 1996; 
Percival et al. 2000) have concluded that while RLQs are 
found exclusively in giant elliptical galaxies, radio-quiet quasars (RQQ) 
reside in both elliptical and spiral (disc dominated) galaxies. It has been 
suggested that the morphological type may depend on the power of the quasar, 
with the most luminous quasars found only in spheroidal host galaxies 
(Taylor et al. 1996). 

Since there is not a homogeneous and  large set of HST observations for
 RLQs,  I have constructed a sample of objects  
 from merging three different subsets 
(Bahcall et al. 1997; Boyce et al. 1998; Dunlop et al. 2001).
Since the subsets have statistically indistinguishable host
luminosity distributions (Treves et al 2001) we have merged these subsamples 
have been merged. 
The combined data set consists therefore of 18 objects with redshift in 
the range 0.158$<z<$0.389, $<z>$=0.26$\pm$0.07 and $<M_{R}>=-24.04\pm0.4$ 
(see also Falomo et al 2002)
As in the case of BLL  an elliptical model  is always a good representation 
for the host galaxies. 
Although based on small (but homogeneous) samples the comparison between 
BL Lacs and RLQs suggests that 
the latter are hosted in galaxies that are systematically  
more luminous by $\sim$0.5 magnitudes. 
High power nuclear activity like that observed in the RLQ sample appear 
therefore to occur only in the most luminous 
and massive galaxies and it is therefore a rare event.

\section{Super massive Black Holes and  host galaxies properties}
There is a large consensus about the existence of supermassive black
holes (SBHs) at the center of nearby inactive galaxies as well as in
the nuclei of active galaxies and quasars (see e.g. Ferrarese 2002 for a recent
review).  A large body of data, in particular based on high
resolution HST observations, is now available 
(see e.g. Kormendy \& Gebhardt 2001) to support the presence of such
massive BH using different techniques.

SBHs play an important role in the formation
and evolution of massive spheroids and are also a key component for the
development of the nuclear activity. In spite of this apparently
ubiquitous presence of SBH in galaxies our understanding on how the
galaxies and their central BHs are linked in the process of formation
of the observed structures is still poorly understood (Silk \& Rees 1998; 
Kauffmann and Helmet 2000; Adams et al. 2001).

From the observational point of view it was shown that ($ {M}_{BH}$) 
is correlated with the bulge mass 
($ {M}_{bulge}$) component of the host galaxy which is translated
into a relationship between $ {M}_{BH}$ and bulge luminosity
$L_{bulge}$ (Magorrian et al. 1998;Kormendy \& Gebhardt 2001 ) and 
between $ {M}_{BH}$ and the velocity dispersion
($\sigma$) (Ferrarese \& Merritt 2000; \cite{FM}; Gebhardt et al. 2000 ).  
These relationships 
are based on a small number ($\sim$40)  of nearby galaxies for which
direct dynamical measurements of $  M_{BH}$ have been secured.
On the other hand, although these empirical relationships have a 
scatter of $\sim$0.4 dex, they offer a new tool
for evaluating $ {M}_{BH}$ of AGN provided that bulge
luminosities and/or velocity dispersion be measured (e.g. McLure \& Dunlop 2002
).

Bettoni et al (2002) have derived the relations between $ {M}_{BH}$ and
$L_{bulge}$ and $\sigma$  using a sample of 20 E-type galaxies  in
the Kormendy \& Gebhardt (2001) galaxy list with measured BH
masses. The two relations are:

\begin{equation}
Log( {M}_{BH}/ {M}_{\odot})=-0.50\times M_{R}-2.97
\label{eq1}
\end{equation}

\begin{equation}
Log( {M}_{BH}/ {M}_\odot) = 4.55\times Log~(\sigma) -2.27,
\label{eq2}
\end{equation}


The two relations were used to evaluate the mass of the BH in RGs, 
BL Lacs and RLQs samples discussed in the previous sections. 
For the sample of 73 RGs  a mean value (using eq. 2) 
$<Log(  M_{BH})>$=8.66$\pm$0.45 is found and a slighter higher value 
if host galaxy luminosities (eq. 1) are used: $<Log(  M_{BH})>$=8.94$\pm$0.37.
The reason for this ($\sim$ a factor of 2) systematic difference  
is not well understood but it needs 
to be taken into account when comparing BH masses using different methods  
(see also discussion in Bettoni et al 2002).

 Only for a small number of BL Lac objects the stellar velocity dispersion of 
the host galaxy has been measured (Falomo Kotilainen and Treves 2002; 
see also Kotilainen et al at this conference ).
The BH mass of 7 BL Lacs derived from measurements of $\sigma$ is 
$<Log(  M_{BH})>$=8.62$\pm$0.23 while using the host galaxy luminosity 
of the  sample of 57 objects at z $<$  0.5 
imaged by HST a slighter higher value of BH mass is found 
$<Log(  M_{BH})>$=8.76$\pm$0.25 (see also Falomo Carangelo \& Treves 2002).

For RLQs no measurements of the stellar velocity dispersion are available 
therefore M$_{BH}$ can be derived only from the luminosity of the 
host galaxy. Again if we consider only objects imaged by HST 
the derived average BH mass is  $<Log(  M_{BH})>$=9.05$\pm$0.20. 
The distributions of BH masses for the three 
subsamples are compared in Figure 2.

\begin{figure}
\plottwo{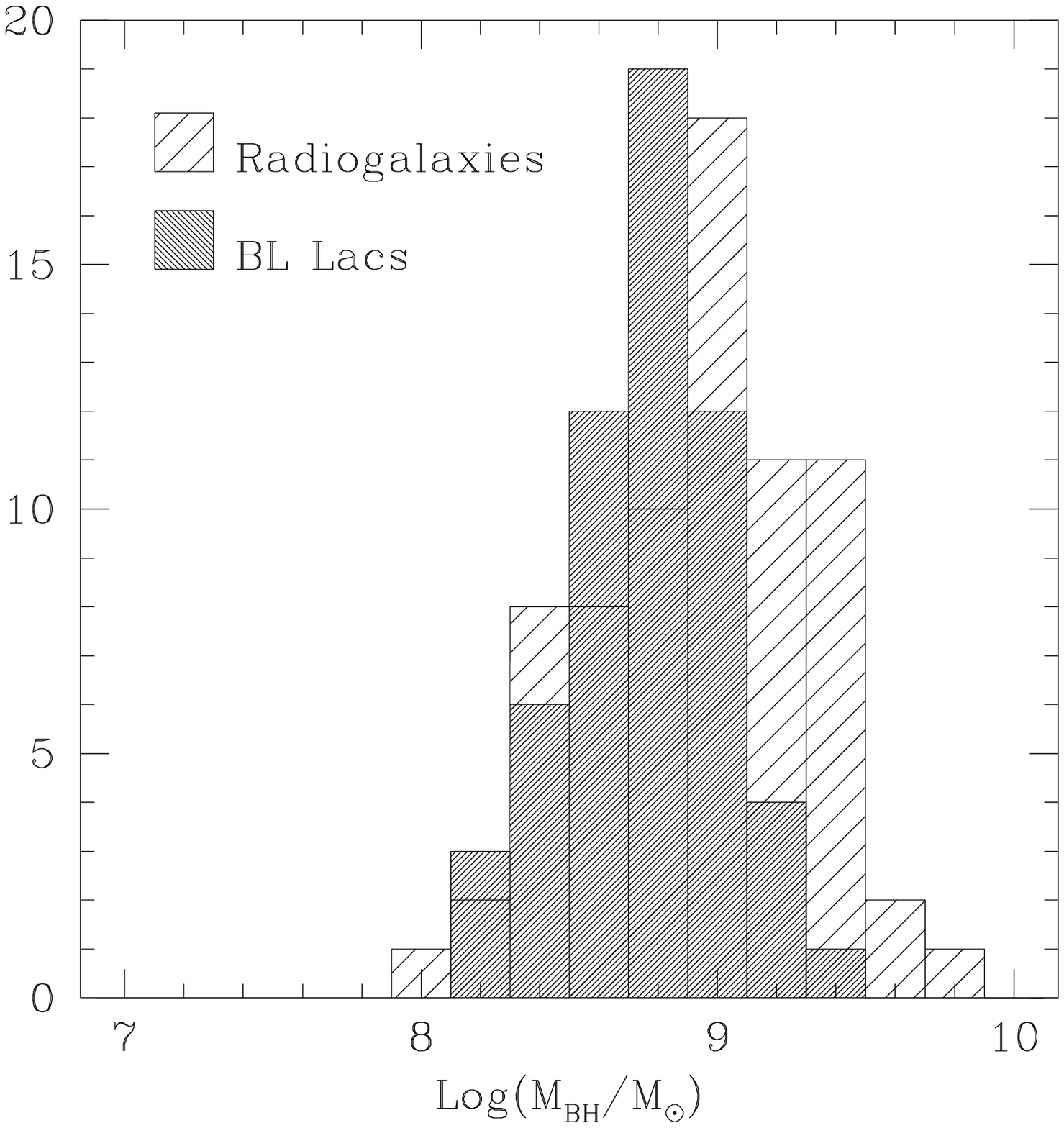}{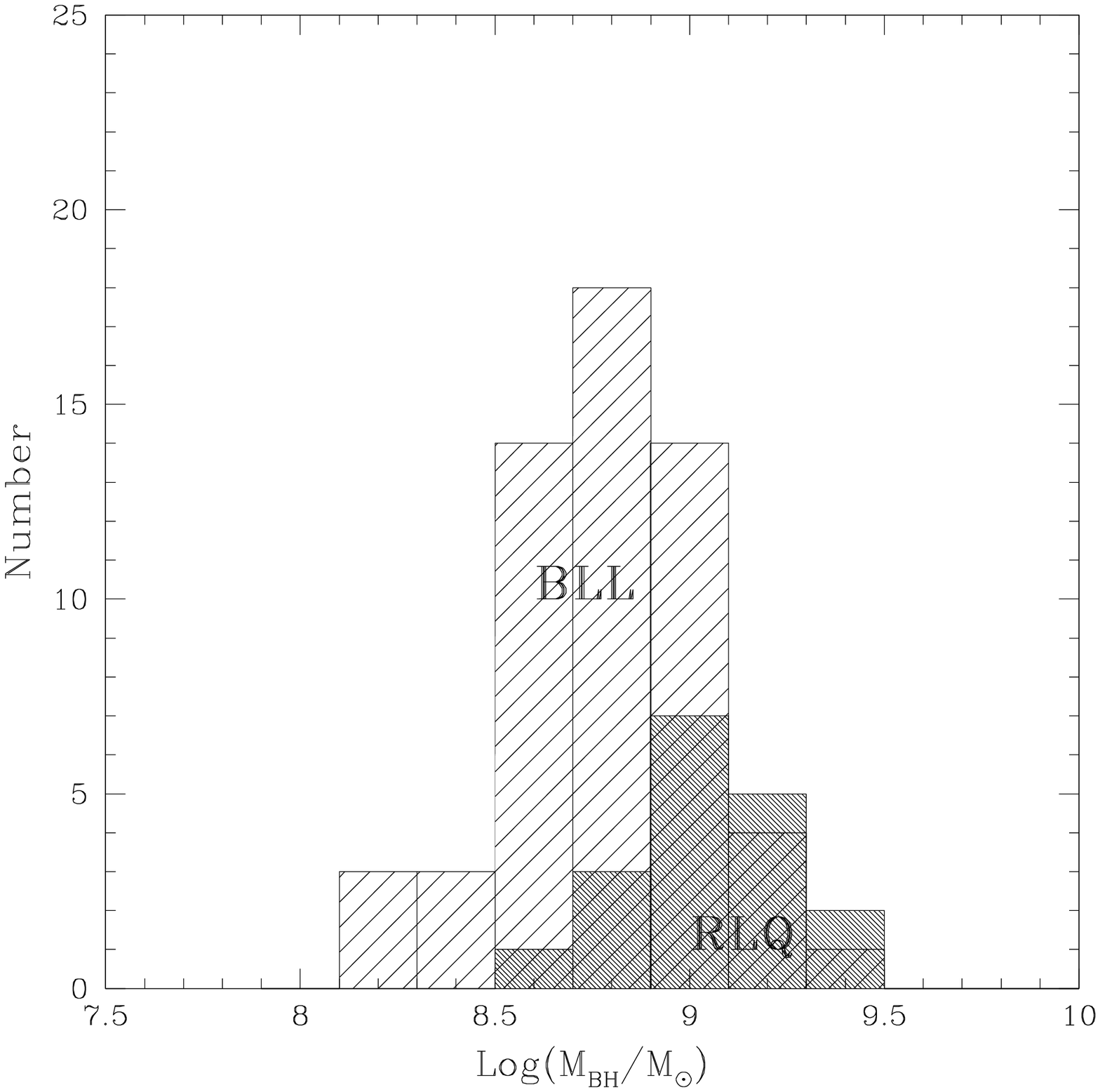}
\caption{Left: Black Hole mass distribution for radiogalaxies (RG) and BL Lacs 
(BLL) using the relationship (see eq. 1 in the text) 
between M$_{BH}$ and M$_R$(host). 
Right: Same as left panel but for RLQs and BLLs .}
\end{figure}

It turns out thus that 
within a factor of two RGs, BLL and RLQs have similar  BH masses but 
their total intrinsic nuclear luminosities are remarkably different.
In addition to the higher observed nuclear/host ratio of RLQ with respect 
to BLL 
for the latter a substantial beaming factor is present ($\delta\sim$15 see 
Ghisellini et al. 1998, Capetti \& Celotti 1999).
Both effects make the intrinsic nuclear luminosities different 
by  a factor $\sim$ 100.
This implies a dramatic difference of the Eddington ratio  
$\xi_{E}=L/L_{E}$ where 
L$_{E}$=1.25$\times$10$^{38}\times$(M$_{BH}$/M$_{\odot}$) erg s$^{-1}$ 
(see also O'Dowd et al 2001; Treves et al 2001).
Basing on the estimated total QSO luminosity 
of L$\sim 3\times$ 10$^{12} L_{\odot}$ (e.g. Elvis et al. 1994) 
and assuming  BH masses of 1x 10$^{9}$ M$_{\odot}$, it is found that 
 RLQ may be emitting at rates of 10\% or higher than their 
Eddington power, while BLL are always  emitting at regimes 
that are much lower than L$_{E}$. 
Given the similarity of  BH masses 
the key  parameters distinguishing RLQs from BLL 
should be considered the accretion rate $\dot{M}$ 
(see also Cavaliere and D'Elia 2002), 
and the jet beaming parameter $\delta$, rather than the black hole mass.

\section{Host galaxies of radio loud active nuclei at high redshift}
A good knowledge of quasar host galaxies is essentially limited
to z $<$ 1 and it  has enabled only a preliminary
insight into the cosmological (z--dependent) quasar--host galaxy
connection. This evolution should become much clearer in the redshift
interval 1 $<$ z $<$ 3, since z$\sim$2 is close to the epoch of the
most vigorous nuclear activity. The observed  similarity of the cosmic quasar
evolution with the rate of galaxy formation (e.g. Franceschini et al
1999) may represent the overall effect of a fundamental link between
massive galaxies and their nuclei, that has driven their formation
history.   

Due to the increasing   difficulty of detecting quasar hosts at high redshift 
only a few studies of RLQ hosts at z $>$ 1 have been conducted so far. 
Using 4m class telescopes (e.g.  Lehnert et al. 1992, 1999;
 Hutchings et al. 1999) the observations suggest RLQs 
are hosted in very luminous galaxies with possibly high star formation rates.
 On the other hand similar studies of RQQs at z$\sim$2.5 (e.g. Lowenthal et
al. 1995) were unable to resolve the hosts.

More recently deep, high spatial resolution  NIR
images of high redshift RLQs obtained either 
using large ground based telescopes (Falomo et al 2002) or HST 
(Kukula et al 2001) show a modest increase with redshift of host luminosity, 
which is consistent with that expected 
from simple passive evolution of massive spheroids.
The luminosities of the high $z$ RLQs hosts (M$_H \sim$ --27.5) are very
similar to those of high redshift 6C RG (Eales et al. (1997). 
The scenario appears, however,  different for RQQ hosts 
since HST and NICMOS imaging of 5 (z $\sim$2) RQQs (
Ridgway et al 2000) indicates their host are 1-2 less luminous.  
than those found in RLQs.

There is also evidence  that the systematic
difference of host luminosity between RLQs and RQQs, already noted at
low redshift (e.g. Bahcall et al. 1997 ), is
more significant at higher redshift. This suggest a
different formation and/or evolutionary history of the two types of
AGN depending on whether or not they can develop radio emission.

Models of galaxy formation and evolution based on hierarchical
clustering (e.g. Kauffmann \& H\"ahnelt 2000) predict progressively
less luminous host galaxies for quasars at high redshift. This seems
to be in reasonable agreement with the observations of high $z$ RQQ hosts
(Ridgway et al. 2000; Hutchings 1995), which may still be undergoing
major mergers to evolve into the low redshift giant ellipticals. 
This scenario,  contrasts
with the results for high $z$ RLQs.
Note, however, that the available data set of the properties of 
high redshift hosts galaxies is still very scanty and further high quality 
observations 
extended up to z $\sim$ 3 are needed to properly assess the above points.
\section{Conclusions and future perspectives}
I have shown that at low redshift (z $<$ 0.5)the galaxies hosting 
radio sources exhibit very similar properties. 
In all classes considered (RG, BL Lacs and RLQs) the hosts are 
luminous galaxies dominated by the spheroidal component. 
These objects seem to follow the behaviour of massive spheroids 
that are well formed at z $>$ 2 and then undergo passive evolution. 
 
Assuming that the relationship between galaxy mass 
(luminosity or velocity dispersion) 
and central black hole mass, found for nearby early type galaxies, 
holds also for these 
more distant active galaxies, the similarity of host properties translates into 
a similarity of M$_{BH}$. The differences in the observed nuclear properties 
must therefore be  either in their viewing angle or/and  in the level of 
accretion. Also a different BH spin could play a relevant role.

Further studies of the host galaxies and their nuclei require to extend 
the analysis to higher redshift. In particular it is important 
to explore, with a statistical data set, QSO hosts up to z $\sim$ 3 in order to 
assess the host evolution around the peak of QSO activity. 
To pursue this goal it is imperative to gather observations 
with high spatial resolution and great efficiency in order to detect the 
faint  nebulosity surrounding the bright nuclei.
These requirements appear well matched by next generation  instrumentation,  
in the near IR, that make use of adaptive optics at large ground based 
telescopes or with the future James Webb Space Telescope  in space.
\acknowledgments{I wish to thank D. Bettoni, N. Carangelo, G. Fasano, 
J. Kotilainen, and A. Treves for  comments and usage of unpublished data.}

\end{document}